\documentclass[9pt]{article}
\usepackage{spconf,amsmath,graphicx}
\usepackage{url}
\usepackage{float} 
\usepackage{subfigure}
\usepackage{amssymb}
\usepackage{multirow}
\usepackage[bottom]{footmisc}
\usepackage{hyperref}
\usepackage{amsmath}
\usepackage{algorithmic}
\usepackage{cite}
\title{Injecting Spatial Information for Monaural Speech Enhancement via Knowledge Distillation}
\name{Xinmeng Xu$^{1}$ \quad Weiping Tu$^{1,2,3,*}$ \thanks{$^*$ Corresponding Author.} \quad Yuhong Yang$^{1, 2}$}
\address{$^1$National Engineering Research Center for Multimedia Software, \\School of Computer Science, Wuhan University,  China\\
$^2$Hubei Key Laboratory of Multimedia and Network Communication Engineering,\\ Wuhan University, China\\
$^3$Hubei Luojia Laboratory, China}
\begin{document}
%\ninept
%
\maketitle
\begin{abstract}
Monaural speech enhancement (SE) provides a versatile and cost-effective approach to SE tasks by utilizing recordings from a single microphone. However, the monaural SE lags performance behind multi-channel SE as the monaural SE methods are unable to extract spatial information from one-channel recordings, which greatly limits their application scenarios. To address this issue, we inject spatial information into the monaural SE model and propose a knowledge distillation strategy to enable the monaural SE model to learn binaural speech features from the binaural SE model, which makes monaural SE model possible to reconstruct higher intelligibility and quality enhanced speeches under low signal-to-noise ratio (SNR) conditions. Extensive experiments show that our proposed monaural SE model by injecting spatial information via knowledge distillation achieves favorable performance against other monaural SE models with fewer parameters.
\end{abstract}
\begin{keywords}
Monaural speech enhancement, spatial information injection, binaural speech enhancement, knowledge distillation.
\end{keywords}
\section{Introduction}
\label{sec:intro}
Monaural speech enhancement (SE) aims to estimate a clean speech from single-channel noisy recording, which is simple but effective and has been widely used as a pre-processor for many speech-based applications. This is because the intelligibility and quality of the enhanced speeches will significantly influence the performance of these downstream tasks. However, it still is a challenging task since many important features in the original speech has be seriously occluded or corrupted by noise.

Recently, deep learning (DL) based monaural SE algorithms, facilitated by powerful general-purpose graphics processing units and large amounts of training data, have shown impressive results by improving speech intelligibility in narrow acoustical conditions \cite{ref1, ref2, ref3}. Specifically, DL-based methods formulate SE as a supervised learning problem, in which the discriminative patterns of speech and noise interference can be learned from training data \cite{ref4, ref5}. Although DL-based methods achieve promising results, their performance still is severely limited since more discriminative patterns of speech and noises are hard to extract from one-channel recording \cite{ref6, ref7}.

Obviously, spatial information is an effective cue that can be exploited to further enhance the SE performance \cite{ref8, ref9}. When multiple microphones are available, the spatial information between different channels can be leveraged and thus the capacity of SE can be further boosted. Consequently, the multi-channel SE frameworks produce lower speech distortion than monaural SE and thus enjoy better generalizability in realistic scenarios. However, to train a multi-channel SE model with high generalization capability, a large number of such noisy/clean speech pairs are needed to sufficiently cover the variations on spatial distribution features of audio signals \cite{ref10}. Fulfilling such a demanding requirement on training samples may be costly and sometimes challenging.

\begin{figure*}[t]
  \centering
  \includegraphics[width=0.95\linewidth]{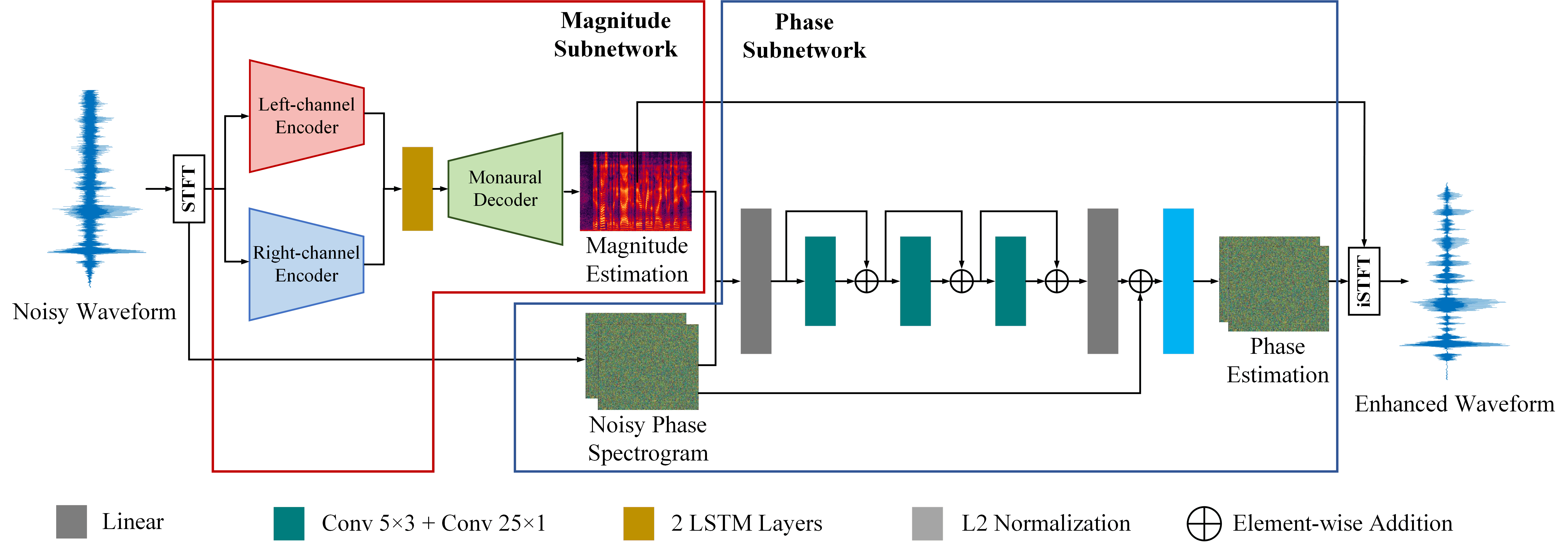}
  \caption{Block diagram of proposed spatial-information injection and enhancement network (SJEN), which consists of a magnitude and a phase sub-module.}
  \label{fig:1}
\end{figure*}

To take advantage of methods in monaural and multi-channel SE and avoid their respective drawbacks, we propose to inject spatial information from the binaural SE model into the monaural model and propose a spatial-information injection and enhancement network (SJEN). The SJEN contains a magnitude and a phase sub-module, which is capable of recovering the magnitude and phase information of target speech. For spatial information injection, a knowledge distillation strategy, which consists of teacher-student and bad student-student frameworks, is specially designed. Specifically, the teacher model, i.e., MHCA-CRN is a binaural SE model under channel-wise encoding structure \cite{ref11}, and the bad student model, i.e., SJEN without knowledge distillation, provides positive and negative knowledge to boost SJEN to learn more accurate binaural speech features from teacher model and to reject features from the bad-student model. Consequently, the monaural SE model uses knowledge of spatial information provided by the teacher model, MHCA-CRN, to realize the automatic generating of spatial features.

\section{Model Description}

In this paper, we propose an efficient spatial-information injection and enhancement network (SJEN) for monaural SE. Meanwhile, we present a novel knowledge distillation strategy to automatically learn spatial features from MHCA-CRN thereby further improving the monaural SE performance.

\subsection{SJEN}

As shown in Figure~\ref{fig:1}, the proposed SJEN takes the acoustic representation extracted from noisy raw waveform using short-time Fourier transform (STFT), which generates magnitude and phase spectrograms, as input. In addition, the SJEN consists of two modules: a magnitude sub-module and a phase sub-module, for producing magnitude and phase estimation. Finally, we transform these estimations back to the enhanced waveform.

\textbf{Magnitude sub-module.} The magnitude sub-module follows the framework of MHCA-CRN \cite{ref11}. Specifically, the two encoders are separately adopted to process speech features of the left and right channels.  Each encoder contains several stacked 2-D convolutional layers, each of which is followed by batch normalization and exponential linear unit (ELU). Then the encoded features of both channels are concatenated and fed to two long-short-term memory (LSTM) blocks for aggregating temporal contexts. The decoder is the mirror representation of each encoder except all the convolution layers are replaced with deconvolutional layers. 

\textbf{Phase sub-module.} Inspired by \cite{ref12}, we design a phase sub-module for phase enhancement by inducing the correlation between magnitude and phase spectrograms. Therefore, the phase sub-module takes the magnitude estimation produced by the magnitude sub-module and noisy phase as input. These two inputs are fused together through linear projection and concatenation over channels \cite{ref13} and then processed by three residual blocks, each of which contains two 2D convolutional layers, one uses a kernel size of $5\times 3$, other uses a kernel size of $25 \times 1$. The phase residual is formed by projecting the result and is added to the noisy phase.  The clean phase prediction is finally obtained by global layer normalization (gLN).

\begin{figure*}[t]
  \centering
  \includegraphics[width=0.95\linewidth]{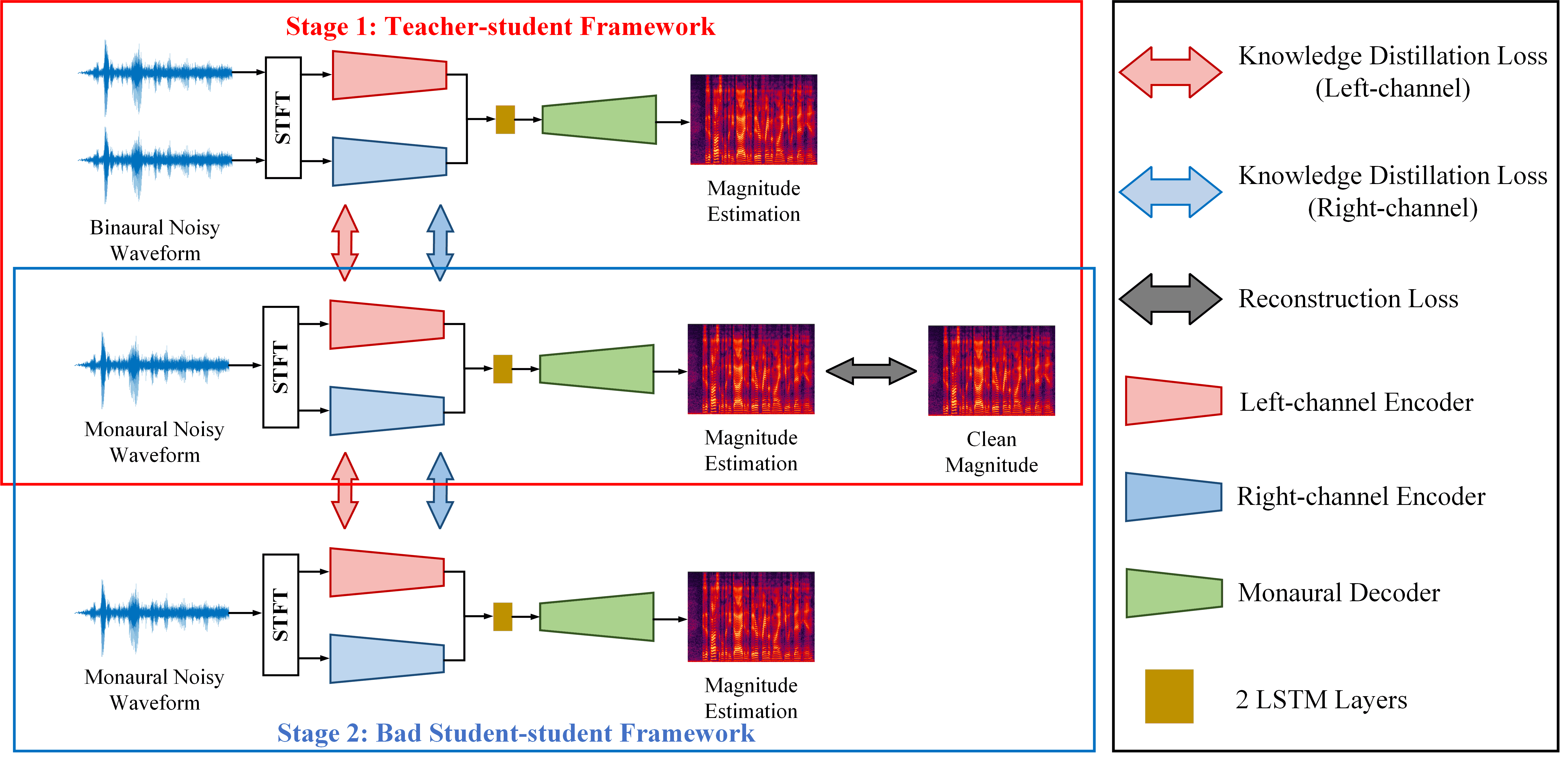}
  \caption{Block diagram of proposed spatial-information injection and enhancement network (SJEN), which consists of a magnitude and a phase sub-module.}
  \label{fig:2}
\end{figure*}

\subsection{Knowledge Distillation}

As shown in Figure~\ref{fig:2}, the proposed knowledge distillation strategy consists of a teacher-student framework and a bad student-student framework for learning positive and negative knowledge from the teacher model and itself.

\textbf{Teacher-student Framework.} Teacher-student framework aims to transfer the knowledge of binaural features from the binaural SE model to the monaural SE model. As shown in Figure~\ref{fig:2} (Top), the framework contains two models, a binaural teacher model, i.e., MHCA-CRN, and a student monaural model, i.e., magnitude sub-module of SJEN. Note that both of these models have the same architecture and take a separate encoding framework that keeps the binaural features for knowledge transfer, but the input of these models are different, i.e., the teacher model takes binaural noisy speech and the student model takes monaural noisy speech. We hope that the features extracted by each encoder layer of the student model can be as similar as possible to the features extracted by each encoder layer of the teacher model. To achieve this, we adopt knowledge distillation loss between the features from the encoder layers of these two models to enable the student model can learn more accurate binaural speech features from the teacher.

\textbf{Bad Student-student Framework.} To ensure that the student model efficiently learns knowledge from the teacher model, we provide a bad student-student framework to provide a negative sample for boosting the learning process. As shown in Figure~\ref{fig:2} (Bottom), the framework also trains two models. It is worth noting that these two models are exactly the same, and the bad student model provides monaural speech features that we treated as negative samples. We hope that the features from the student model can be pulled closer to the feature from the teacher and pushed far away from the feature of bad student. To achieve this, we maximize the knowledge distillation loss between features from encoder layers of the student and bad student model.

\subsection{Loss Function}

\quad \quad \textbf{Reconstruction Loss.} Following previous works, The magnitude sub-module is trained by minimizing L2 loss to measure the difference between the magnitude estimation and the ground-truth. The phase sub-module is trained by maximizing the cosine similarity between the phase prediction and ground truth \cite{ref13}:
\begin{equation}
\begin{aligned}
     \mathcal{L}_{RL} &= ||\hat{M}-M||_2 \\ &- \alpha \frac{1}{TF}\sum_{t, f}M(t, f)<\hat{P}(t, f), P(t, f)>,
\end{aligned}
\end{equation}
where $\hat{M}$, $M$, $\hat{P}$, and $P$ represent magnitude estimation, ground-truth magnitude spectrogram, phase estimation, and ground-truth phase spectrogram respectively.

\textbf{Knowledge Distillation Loss.} To ensure that the proposed strategy can truly realize knowledge transfer, the knowledge distillation loss is proposed, which is essentially a feature-matching function. We define the knowledge distillation loss for teacher-student framework, $\mathcal{L}_{KDL}^{TS}$, and bad student-student framework, $\mathcal{L}_{KDL}^{BS}$, as
\begin{align}
    \mathcal{L}_{KDL}^{TS} &= \sum\nolimits_{i=1}^5||(S_L^i-T_L^i) + (S_R^i-T_R^i)||_1, \\
    \mathcal{L}_{KDL}^{BS} &= \sum\nolimits_{i=1}^5||(S_L^i-B_L^i) + (S_R^i-B_R^i)||_1,
\end{align}
where $S_L^i$, $T_L^i$, and $B_L^i$ denote the output feature of the $i^{th}$ encoder layer for the left channel in the student, teacher, and bad student models, respectively. Meanwhile, we hope that $\mathcal{L}_{KDL}^{TS}$ can be as smaller as possible and $\mathcal{L}_{KDL}^{BS}$ can be as larger as possible, thus the total knowledge distillation loss is
\begin{equation}
    \mathcal{L}_{KDL}^{Total} = \mathcal{L}_{KDL}^{TS}/\mathcal{L}_{KDL}^{BS}.
\end{equation}

Finally, the total loss of whole process is defined as
\begin{equation}
    \mathcal{L}_{Total}=\mathcal{L}_{RL} + \beta \mathcal{L}_{KDL}^{Total}.
\end{equation}

\renewcommand{\arraystretch}{0.90}
\begin{table*}[]
\centering
\caption{PESQ and STOI comparison for different models. A higher score means better performance where the \textbf{BOLD} text indicates the best performance for each metric.}
\begin{tabular}{l|cc|cc|cc|cc}
\hline
Test SNR                           & \multicolumn{2}{c|}{-5 dB}     & \multicolumn{2}{c|}{0 dB}      & \multicolumn{2}{c|}{5 dB}      & \multicolumn{2}{c}{10 dB}      \\ \hline
Metric                             & STOI (\%)      & PESQ          & STOI (\%)      & PESQ          & STOI (\%)      & PESQ          & STOI (\%)      & PESQ          \\ \hline
Unprocess                          & 57.71          & 1.51          & 71.02          & 1.82          & 82.53          & 2.12          & 90.41          & 2.44          \\ \hline
CRN \cite{ref5}                                & 77.54          & 1.74          & 86.34          & 2.36          & 94.35          & 2.68          & 96.49          & 3.04          \\
SA-TCN \cite{ref17}                            & 82.95          & 2.16          & 89.17          & 2.61          & 95.19          & 2.93          & 96.56          & 3.16          \\
PHASEN \cite{ref12}                            & 83.84          & 2.26          & 90.23          & 2.74          & 95.56          & 3.18          & 96.43          & 3.24          \\ \hline
SJEN (Proposed)                    & \textbf{85.32} & \textbf{2.30} & \textbf{91.06} & \textbf{2.83} & \textbf{95.84} & \textbf{3.20} & \textbf{96.79} & \textbf{3.28} \\
-w/o phase sub-module              & 83.99          & 2.24          & 90.32          & 2.75          & 95.53          & 3.10          & 96.38          & 3.25          \\
-w/o bad student model             & 84.83          & 2.27          & 90.86          & 2.78          & 95.78          & 3.14          & 96.64          & 3.26          \\
-w/o teacher and bad student model & 77.68          & 1.86          & 87.10          & 2.41          & 94.97          & 2.74          & 96.36          & 3.09          \\ \hline
\end{tabular}
\label{tab:1}
\end{table*}

\section{Experimental Setup}
\subsection{Datasets}
We use WSJ0 \cite{ref14} as clean speech dataset, which includes 7138 utterances from 83 speakers (42 males and 41 females). The noisy speech is a mixture of the above clean speech segments with the noise dataset NOISEX-92 \cite{ref15}. The binaural room impulse responses (BRIRs) for binaural speech simulation using the Multi-Channel Impulse Responses (MCIRs) Database. 

The clean speech scaling process is $\hat{x}(n)=\nu x(n)$ where $\nu =10^{\epsilon/20}/\sigma_x$ with $\epsilon$ being a value randomly selected from $-35:1:-15$ dB and $\sigma_x = \sqrt{E[x^2(n)]}$ with $E[\cdot]$ denoting mathematical expectation. The noise is rescaled as $\hat{v}(n)=\vartheta v(n)$, where $\vartheta=10^{\frac{-\texttt{SNR}}{10}\sigma^2_{\hat{x}}/\sigma^2_v}$, $\sigma^2_{\hat{x}}=E[\hat{x}^2(n)]$, $\sigma^2_v=E[v^2(n)]$, and SNR is randomly chosen from $-15:1:-15$ dB \cite{ref16}. The data for teacher model is simulated as
\begin{align}
    &y_L(n) =\hat{x}(n)*h_{x,L}(n)+\hat{v}(n)*h_{v,L}(n),\\
    &y_R(n) =\hat{x}(n)*h_{x,R}(n)+\hat{v}(n)*h_{v,R}(n),
\end{align}
where $h_{x,L}(n)$, $h_{x,R}(n)$, $h_{v,L}(n)$, and $h_{v,R}(n)$, are, respectively, the acoustic impulse responses from MCIRs database. The data for student and bad student model is simulated as
\begin{equation}
    y(n) = \hat{x}(n) + \hat{v}(n).
\end{equation}

A total of 40 different noise conditions are considered in the training set and 20 different conditions are considered in the test set. Finally, the training and test set contains 11572 and 824 noisy clean speech pairs, respectively.

\subsection{Baselines and training details}
We trained 3 recent monaural SE models for performance comparison, (1) \textbf{CRN} \cite{ref5}: the convolutional recurrent network for SE, (2) \textbf{SA-TCN} \cite{ref17}: the multi-stage self-attentive temporal convolutional networks for SE, (3) \textbf{PHASEN} \cite{ref12}: a phase-and-harmonics-aware SE network.

For the training step, all models are trained with Adam optimizer for stochastic gradient descent (SGD) based optimization. The learning rate is set to 0.001. All training samples are zero-padded to have the same number of time steps as the longest sample.

\section{Results and Analysis}

Comprehensive comparison among alternative models, are shown in Table~\ref{tab:1}, in terms of perceptual evaluation speech quality (PESQ) and short-term objective intelligibility (STOI).

\renewcommand{\arraystretch}{0.95}
\begin{table}[]
\centering
\caption{Quantitative results for different models.}
\begin{tabular}{l|ccc}
\hline
Models          & FLOPs (G) & Param. (M) & RTF  \\ \hline
CRN \cite{ref5}            & 1.16      & 15.83       & 0.12 \\
SA-TCN \cite{ref17}         & 8.85      & 8.54       & 0.35 \\
PHASEN \cite{ref12}         & 2.85      & 6.41       & 0.22 \\
SJEN (Proposed) & 1.14      & 1.56       & 0.14 \\ \hline
\end{tabular}
\label{tab:2}
\end{table}

The proposed SJEN consistently outperforms CRN, SA-TCN, and PHASEN almost in all conditions. Next, we observe that the phase estimated by the phase sub-module of SJEN performs than mixed phase, i.e., SJEN vs. SJEN-w/o phase sub-module. Finally, according to Table~\ref{tab:1}, we observe that the knowledge transferred from teacher model and the bad student model does promote the network performance significantly. The knowledge of binaural speech features from teacher model improves $7.20\%$ and 0.18 on STOI and PESQ, respectively at -5 dB SNR condition. In addition, the negative knowledge provided by bad student model improves $0.50\%$ and 0.04 on STOI and PESQ, respectively at -5 dB SNR condition. 

The quantitative results are presented in Table~\ref{tab:2}. Besides STOI and PESQ evaluation metrics, we also report the floating-point operations (FLOPs), the number of trainable parameters, and real-time factor (RTF). Obviously, SJEN is significantly better than the SA-TCN and PHASEN with fewer FLOPs, number of and RTF.

\section{Conclusion}
In this paper, we propose a spatial-information injection and enhancement network (SJEN) for monaural SE, which improves noise reduction by transferring binaural speech features from binaural SE model to monaural SE model. In addition, we propose a knowledge distillation strategy that consists of teacher-student and bad student-student framework for providing positive and negative knowledge to boost knowledge transfer. Experimental results show that our model achieves superior performance and outperform recent models with fewer parameters.

\vfill\pagebreak

\bibliographystyle{IEEEbib}
\bibliography{refs}

\end{document}